\documentclass[]{spie}  %>>> use for US letter paper
\usepackage[]{graphicx}
\usepackage{graphicx}
\usepackage{amsmath}
\usepackage{amssymb} %squares and such
\usepackage{mathrsfs} %mathscr font
\usepackage{acronym}

\DeclareGraphicsExtensions{.jpg,.pdf,.png,.eps,.ps}

\newcommand{\sinx}{${\rm Si}_3{\rm N}_x$}
\newcommand{\siotwo}{${\rm SiO}_2$}

\acrodef{FTS}{fourier transform spectroscopy}
\acrodef{CMB}{cosmic microwave background}
\acrodef{NEP}{noise equivalent power}
\acrodef{MKID}{microwave kinetic inductance detector}
\acrodef{KID}{kinetic inductance detector}
\acrodef{LEKID}{lumped-element KID}
\acrodef{sinx}[\sinx]{amorphous silicon-nitride}
\acrodef{TiN}{titanium nitride}
\acrodef{JPL}{the Jet Propulsion Laboratory}
\acrodef{MDL}{the Microdevices Laboratory}
\acrodef{TLS}{two-level system}
\acrodef{CPW}{coplanar waveguide}
\acrodef{SOI}{silicon-on-insulator}
\acrodef{VNA}{vector network analyzer}
\acrodef{IDC}{interdigitated capacitor}
\acrodef{CCAT}{Cerro Chajnantor Atacama Telescope}

\title{MKID development for SuperSpec: an on-chip, mm-wave, filter-bank spectrometer}

\author{E. Shirokoff\supit{a},
P. S. Barry\supit{b},
C. M. Bradford\supit{c}, 
G. Chattopadhyay\supit{c}, 
P. Day\supit{c}, 
S. Doyle\supit{b}, 
S. Hailey-Dunsheath\supit{a}, 
M. I. Hollister\supit{a,c},
A. Kov\'{a}cs\supit{a,d}, 
C. McKenney\supit{a}, 
H. G. Leduc\supit{c}, 
N. Llombart\supit{e}, 
D. P. Marrone\supit{f}, 
P. Mauskopf\supit{b,g}, 
R. O'Brient\supit{a,c}, 
S. Padin\supit{a}, 
T. Reck\supit{c},
L. J. Swenson\supit{a,c},
J. Zmuidzinas\supit{a,c}
\skiplinehalf
\supit{a}California Institute of Technology, Pasadena, CA, U.S.A.\\
\supit{b}School of Physics \& Astronomy, Cardiff University, Cardiff, Wales, U.K.\\
\supit{c}Jet Propulsion Laboratory, Pasadena, CA, U.S.A.\\
\supit{d}Univ. of Minnesota, Twin Cities, MN, U.S.A. \\
\supit{e}Complutense University of Madrid, Spain; \\
\supit{f}Steward Observatory \& University of Arizona, Tucson, AZ, U.S.A; \\
\supit{g}Arizona State University, Tempe, AZ, U.S.A.\\
}

\authorinfo{Direct correspondence to E. Shirokoff: erik.shirokoff@caltech.edu}
\begin{document} 
  \maketitle 

%%%%%%%%%%%%%%%%%%%%%%%%%%%%%%%%%%%%%%%%%%%%%%%%%%%%%%%%%%%%% 
\begin{abstract}
SuperSpec is an ultra-compact spectrometer-on-a-chip for 
millimeter and submillimeter wavelength astronomy.  
Its very small size, wide spectral bandwidth, and highly multiplexed readout will enable construction of powerful multibeam spectrometers for high-redshift observations.  
The spectrometer consists of a horn-coupled microstrip feedline, a bank of narrow-band 
superconducting resonator filters that provide spectral selectivity, and 
\acp{KID} that detect the power admitted 
by each filter resonator.  
The design is realized using thin-film lithographic structures on a silicon wafer.  
The mm-wave microstrip feedline and spectral filters of the first prototype are designed to operate in the 
band from 195-310 GHz and are fabricated from niobium with at $T_c$ of $9.2\,K.$  
The KIDs are designed to operate at hundreds of MHz and are fabricated from 
titanium nitride with a $T_c$ of $\sim 2\,$K. Radiation incident on the horn 
travels along the mm-wave microstrip, passes through the 
frequency-selective filter, and is finally absorbed by the corresponding 
KID where it causes a measurable shift in the resonant frequency. 
In this proceedings, we present the design of the \acp{KID} employed in 
SuperSpec and the results of initial laboratory testing of a prototype device.  
We will also briefly describe the ongoing development of a demonstration instrument 
that will consist of two 500-channel, R=700 spectrometers, one operating in the 1-mm 
atmospheric window and the other covering the 650 and 850 micron bands.
\end{abstract}

%>>>> Include a list of keywords after the abstract 

\keywords{kinetic inductance detector, MKID, resonator, titanium nitride, mm-wavelength, spectroscopy}

%%%%%%%%%%%%%%%%%%%%%%%%%%%%%%%%%%%%%%%%%%%%%%%%%%%%%%%%%%%%%
\section{INTRODUCTION}
\label{sec:intro}  % \label{} allows reference to this section
The epoch of reionization, and the birth and subsequent growth of galaxies and clusters in the first half of the Universe's history are key topics in modern astrophysics.  At present, the bulk of our information about this important epoch comes from studies in the rest-frame ultraviolet. Yet measurement of the cosmic far-IR background indicates that in aggregate, half the energy released by stars, star formation, and accretion through the Universe's history has been absorbed and reradiated by dust and gas at mm and submm wavelengths.

Spectroscopic surveys at millimeter wavelengths, using a multi-pixel spectrometer such as we describe here on a large telescope, are uniquely poised to access the high-redshift Universe, both through the measurement of individual galaxies, and via statistical studies in wide-field tomography.   In particular, the $157.7\mu$m [CII] transition is typically the brightest spectral feature in dusty galaxies.  It carries $\sim 0.1\%$ of the total luminosity, and promises to be a powerful probe of galaxies beyond redshift 3.0 where it is redshifted into the atmospheric transmission windows at wavelengths $\ge 600\,\mu$m.

[CII] spectroscopy, combined with dust continuum measurements, allows for the determination of temperature and luminosity, and provides a direct and unbiased measure of the galaxy luminosity function and the history of star formation.  Additional science targets include the evolution of atomic and molecular gas properties via [CII] and CO line surveys of optical catalog targets, and the unique ability to measure galaxy clustering and the galaxy power spectrum ($P(k)$) at high redshifts ($z>4$).

SuperSpec is a novel, ultra-compact spectrometer-on-a-chip for millimeter and submillimeter wavelength astronomy.  Its very small size, wide spectral bandwidth, and highly multiplexed detector readout will enable construction of powerful multibeam spectrometers for high-redshift observations.  This filter-bank spectrometer employs high-density, planar, lithographic fabrication techniques, and easily fabricated and naturally multiplexed \acp{KID}.

During the next year, we will build a  demonstration instrument using the SuperSpec technology.  This camera will include at least two spectrometer pixels, one in the $195$-$310\,$GHz atmospheric band, and one in the $320$-$520\,$GHz band, each with $~600$ channels with a resolution of $\mathcal{R}=700$.  We then aim to apply this technology to a proposed direct-detection wide-band survey spectrometer for the \ac{CCAT} with tens to hundreds of pixels.  The proposed instrument, nominally called {\em X-Spec} will be optimized for measuring the bright atomic fine-structure and molecular rotational transitions from interstellar gas in galaxies.  A 300-pixel, dual-band spectrometer based on this technology would be more than an order of magnitude faster than ALMA for full-band extra-galactic surveys.

\section{MM-WAVELENGTH CIRCUIT}

The \ac{KID}-based design outlined here is a specific example of the filter-bank spectrometer circuit discussed in more detail in Kov\'{a}cs et. al.\cite{Kovac12}. The basic design is similar to other filter-bank spectrometers which appear in the literature\cite{tauber91,galbraith08b,galbraith96}, as well as the contemporary project DESHIMA.\cite{endo12}  In this design, incoming radiation propagates down a transmission line (the feedline) and past a series of $N_c$ tuned resonant filters.  Each filter consists of a section of transmission line with a length of $\lambda_i/2$ where $\lambda_i$ is the propagation wavelength corresponding to channel $i$ with center frequency $f_i$.  The transmission line resonators are coupled to the feedline and to power-detectors with independently adjustable coupling strengths, ie. capacitors, or in the specific example discussed here, proximity coupling between microstrip lines.  The resonator frequencies are arranged monotonically and physically spaced at approximately $\lambda_i/4$ from neighboring channels.  The response of an individual filter channel can be modeled as a dissipative shunt resonator, with a coupling-$Q$ ($Q_{\rm feed}$) to the feedline and a dissipative-$Q$ ($Q_{\rm det}$) associated with the power per cycle deposited in the detector.  The resonator resolution is given by
\begin{equation}
\mathcal{R} = Q^{mm}_{r} = \Big( \frac{1}{Q_{\rm feed}} + \frac{1}{Q_{\rm det}}\Big)^{-1} = \frac{f_i}{\Delta f_i}
\end{equation}

\noindent where $f_i$ and $\Delta f_i$ are the center frequency and width of channel $i$, and $Q^{mm}_r$ is the total resonator $Q$.  Maximum absorption occurs when $Q_{\rm feed} = Q_{\rm det}$.  The full spectrometer is created by starting from the highest frequency channel $f_u$ , then decreasing in frequency according to a geometric progression: $f_u$ , $x\,f_u$ , $x^2\,f_u$, . . . $x^{N_c-1}\,f_u$, where $x < 1$ is the frequency scaling factor, given by
\begin{equation}
x=\exp \Big( - \frac{\ln f_u - \ln f_l}{N_c-1}\Big)
\end{equation}

\noindent and $N_c$ is the total number of channels.  The ratio of the channel spacing to resolution is given by the oversampling factor $\Sigma$, where

\begin{equation}
N_c = \Sigma \mathcal{R} \ln (f_u/f_l).
\end{equation}

\noindent As $\Sigma$ increases, the total in-band absorption efficiency will become larger than 50\%, the maximum value for an isolated single resonator.  For an initial SuperSpec demonstration instrument with 600 channels with $\mathcal{R}=700$ arranged from 195 to $310\,$GHz, $\Sigma = 1.85$ and the total in-band absorption efficiency is approximately 80\%. (Neglecting losses in the feed line, etc.)  

\begin{figure}
   \begin{center}
   \begin{tabular}{c}
   \includegraphics[width=.95\textwidth]{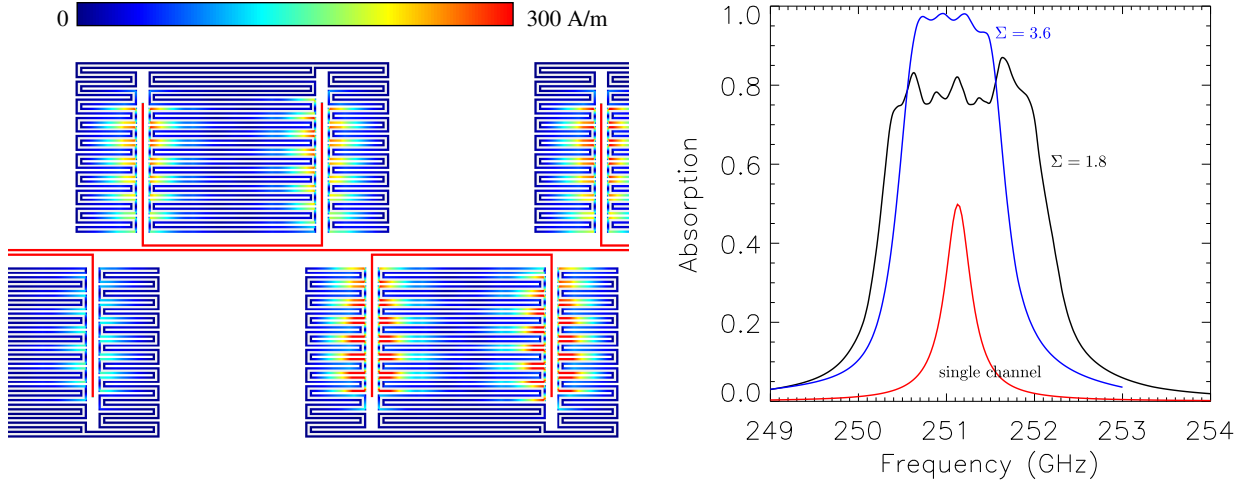}
   \end{tabular}
   \end{center}
   \caption[example] 
   { \label{fig:mm1}
     {\bf (left)} A simulation showing the time-averaged magnitude of the current present in part of an 8-channel SuperSpec filter bank when driven at a specific frequency.  mm-wavelength radiation incident from the left along the central microstrip feedline couple to U-shaped microstrip resonators, which in turn excite currents in the TiN meanders. {\bf (right)}  The total power absorbed for the 8-channel filter bank with two different oversampling factors, with the response of a single isolated channel shown for comparison.}
   \end{figure}

Figure~\ref{fig:mm1} shows the implementation of this concept for the SuperSpec prototype.  The mm-wave circuit design summarized here and its coupling to a metal feedhorn are discussed in more detail in Barry et al.\cite{Barry12}. In this design, the feedline and resonant filters are realized using inverted microstrip.  This microstrip consists of superconducting Nb traces with a $1\,\mu$m width on Si substrate  beneath a $0.5\,\mu$m thick \ac{sinx} dielectric and a Nb ground plane.  This feedline structure provides a $30\,\Omega$ characteristic impedance, suffers negligible radiation loss, allows for adequate proximity coupling between feedlines, and can be readily coupled to the \ac{LEKID} design discussed below without the need for vias or challenging step-coverage solutions.

The U-shaped half wave resonators are fabricated from the same Nb microstrip as the feedline.  The center portion is proximity coupled to the feedline, while the tines couple to a lossy meander made from \ac{TiN}.  Radiation at frequencies above the superconducting gap in the \ac{TiN} film ($T_c\sim 2\,$K) breaks superconducting (Cooper) pairs and generates quasiparticles in the \ac{TiN} \ac{LEKID} inductive meander.
This results in a perturbation of the complex impedance ($\delta \sigma(\omega) = \delta \sigma_1 - j \delta \sigma_2$) producing a measurable change in the dissipation and resonant frequency of the \ac{LEKID}.  
The \ac{LEKID} response is thus a direct measure of the power dissipated in the device.
In the mm-wave circuit, the $20\,$nm thick \ac{TiN} meander can be regarded a resistive $\sim 50\,\Omega/\square$ material.

$Q_{\rm feed}$ and $Q_{\rm det}$ can be adjusted by varying the length of the overlap region and the gaps separating the half wave resonator, the feedline, and the TiN meander.  Starting with an estimate based on the analytic treatment of coupled microstrip lines in Abbosh et al.\cite{abbosh2009}, the final design for the prototype filter bank was based on simulations using Sonnet Software\footnote{http://www.sonnetsoftware.com\/}, a commercial, 3D planar, method-of-moments solver.  Intended values of $f_0$, $Q_{\rm feed}$, $Q_{\rm det}$ are interpolated on a grid of simulation results to obtain design values for $l_0$, $G$, and $G_a$, the resonator length, gap between the resonator and feed, and gap between the resonator and absorber, respectively.  To allow for rapid simulation, infinitely-thin films are used in the model, and an additional correction is then included by hand to account for the improved coupling associated with finite film thickness based upon a sparse set of thick film simulations. 

In practice, the the feed-resonator gap appropriate for our target $\mathcal{R} \sim 700$ channels is the most critical lithographic dimension, which has driven us to maximize the overlap region consistent with the $\lambda/4$ spacing of our highest frequency channels.  The Deep-UV lithography stepper at the \ac{MDL} at \ac{JPL} can readily achieve line widths and feature spacing of $~\sim 1 \pm 0.1 \mu{\rm m}$, which should allow for the construction of well matched filters with any value of $\mathcal{R}$ larger than approximately $200$.  For an $\mathcal{R}=700$ absorber, variations in the Nb linewidth by $0.1\,\mu$m lead to a $\pm 8\%$ change in $\mathcal{R}$ and a negligible change in absorption efficiency.  The thickness of the \ac{sinx} dielectric is the most significant material parameter; the magnitude of this error approaches to our lithographic tolerances at the level of $~3\%$ total thickness variation.  (The circuit is equally sensitive to the dielectric constant of this material, though we expect less wafer-to-wafer variation in this parameter.)  The cumulative effect of less-critical tolerances, such as the TiN linewidth, mask alignment, and TiN thickness and resistivity lead to an additional and largely orthogonal $5\%$ variation in $\mathcal{R}$.

\section{KID design}
\begin{figure}[hbt!]
   \begin{center}
   \begin{tabular}{c}
   \includegraphics[width=.95\textwidth]{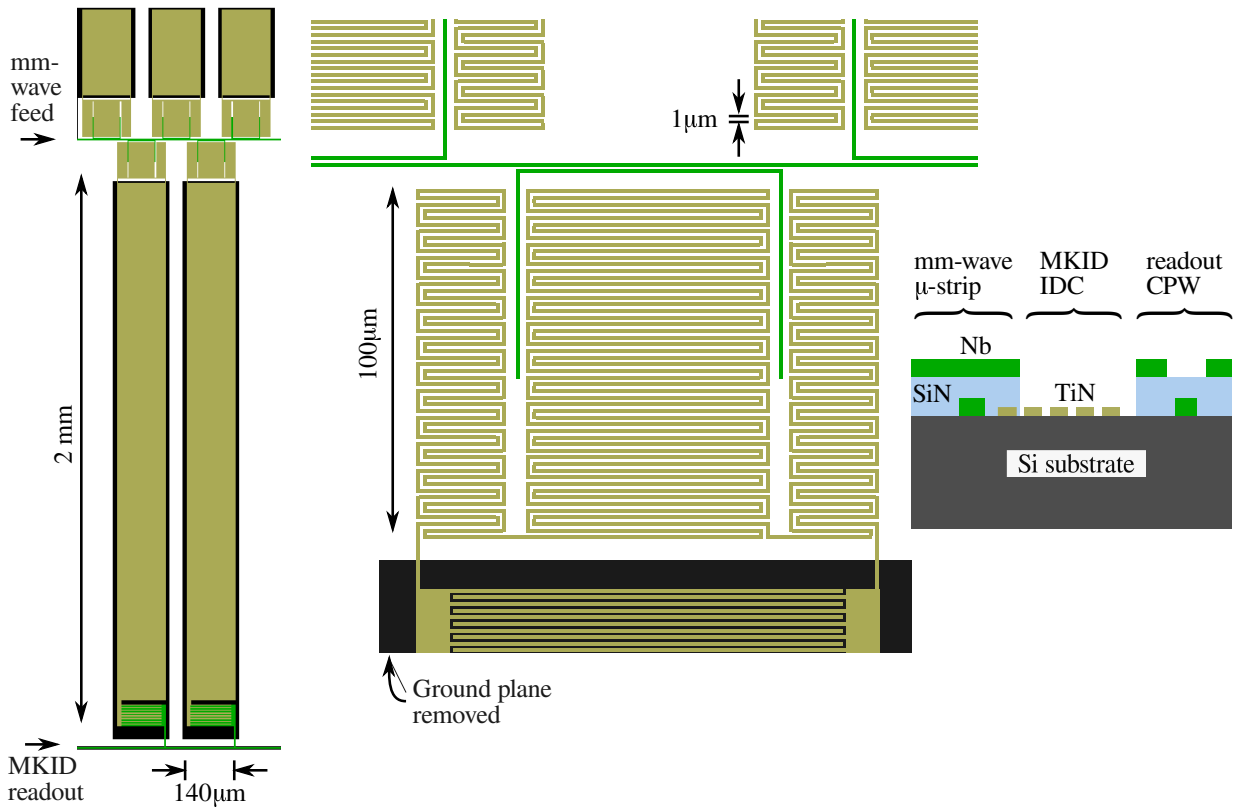}
   \end{tabular}
   \end{center}
   \caption[example] 
   { \label{fig:wholekid01}
     {\bf (center)} The mm-wave portion of a single channel.  Radiation from the left propagates along the green Nb microstrip line and excites the U-shaped half-wave resonator, which in turn couples to the amber TiN meander.  The TiN meander forms the inductive portion of a KID, and is connected to a large IDC. {\bf (left)} A wider view showing several nearby channels.  Below the KID IDC, a second, smaller IDC couples the KID resonator to a readout line, made from bridged Nb CPW.  In the region around the IDCs, shown in black, the ground plane and dielectric have been removed.  {\bf (right)} A cross-section showing the device layers in the region of the mm-wave circuit, the IDC, and the readout line.  Not shown is a thin ${\rm SiO}_2$ protective layer which is deposited between the TiN and the nitride and removed with an HF dip from the region around the IDC.
   }
   \end{figure}

The signal power admitted by each SuperSpec filter resonator is read out using a \ac{LEKID} that consists of an inductive portion made from the TiN meander described previously and a \ac{IDC} made from the same material, as shown in figure~\ref{fig:wholekid01}.  These KIDs make use of the novel properties of \ac{TiN} films: high normal-state resistivity of $\sim 100\,\mu \Omega$-cm and thus a large kinetic inductance fraction, a variable critical temperature which can be adjusted by varying deposition conditions, and low losses suitable for making high-$Q$ resonators.  To estimate material properties for this design we use values from Leduc et. al.\cite{leduc10}

The design of the physical dimensions of the SuperSpec KIDs begins with a multiplexing density specification: we require the ability to multiplex one 600-channel spectrometer within a single octave of readout bandwidth.  We can then determine a minimum required value for the resonator $Q_r$, where 

\begin{equation}
\frac{\Delta f}{f} = \frac{1}{Q_r}=\frac{1}{Q_i} + \frac{1}{Q_c},
\end{equation}

\noindent and $Q_i$ and $Q_c$ are the internal and coupling $Q$s of the resonator.  (Note that the resonators and $Q$ values here all refer to the readout circuit operating at frequencies of a few hundred MHz, rather than the mm-wave filterbank circuit.)

The values of $Q_r$ required to avoid unacceptable losses to collisions between resonators is then calculated using a Monte-Carlo simulation that assumes the following: the resonator targets are uniformly distributed in logarithmic frequency, each resonator at design frequency $f_i$ scatters randomly by an amount $\delta_i$ described by a Gaussian probability distribution with $\sigma_f=\sqrt{<\delta_i/f_i>}$,  and any two resonators whose shifted positions lie within five times the resonator bandwidth are lost.  For plausible values of $\sigma_f$ equal to 0.001 and 0.01, choosing $Q = 10^5$ results in the loss of 2.4\%, and 3.6\% of channels, respectively.  These values approach the most optimistic expectations for fabrication yield, and suggest that we can comfortably choose $Q_r = 10^5.$ 

Although various resonator architectures can, in principle, produce devices with $Q_i>10^5$ and sufficiently low \ac{TLS} noise, we've chosen the conservative approach and use \acp{IDC} patterned on bare, crystal Si substrate; a resonator design which reliably produces high-$Q$ resonators with relatively low \ac{TLS} noise.  To avoid the need for vias and complicated step-coverage, we have therefore chosen the inverted microstrip design for the mm-wave circuit discussed above, which allows the TiN layer deposition to occur as the first processing step. (See figure~\ref{fig:wholekid01}.)

By operating at a readout frequency of a few hundred MHz, we can significantly reduce the cost and complexity of readout hardware and also reduce the effect of \ac{TLS} noise by increasing $\beta(\omega)$, the ratio of the frequency response to the dissipation response of a resonator, which scales as $\beta \propto kT/\omega$.~\cite{zmuidzinas12}  For the prototype device, we've chosen to use the $100-200\,$MHz range.  Future mutli-pixel implementations will likely cover several octaves of bandwidth.

With the required $Q$ for multiplexing and readout frequency specified, the SuperSpec inductor is then designed to meet the following set of requirements.  (1) $T_c$ must be less than $2.6\,$K in order to absorb photons at the $190\,$GHz edge of our observation band.  (2) The internal $Q$ associated with dissipation loss due to optically generated quasiparticles must be compatible with the $Q\sim 10^5$ requirement from multiplexing considerations.  (3) The inductor area should be largely contained with a few dissipation lengths of the region where mm-wave absorption occurs. (4) The operating temperature should be as high as possible, in order to minimize \ac{TLS} noise, while still satisfying the requirement that the density of optically-generated quasiparticles should be significantly larger than the density of thermally generated quasiparticles. (5) The value of the inductance should be maximized, subject to the above constraints, in order to minimize die size.

Using the standard Mattis-Bardeen formulas for the complex conductivity of a superconductor (as summarized in Zmuidzinas et. al.\cite{zmuidzinas12}), and implementing the above conditions, we arrive at a design for the SuperSpec KID with physical dimensions and estimates of physical properties given in table \ref{table:properties}.  This design uses the thinnest TiN film which we expect to be able to deposit with high yield and well-controlled properties, with $T_c=2K$ and an operating temperature of $250\,$mK appropriate for a He-sorption refrigerator.

\begin{table}[h]
\caption{Summary of the expected properties of the SuperSpec detectors for a demonstration instrument} 
\label{table:properties}
\begin{center}
\begin{tabular}{|l|l|} %% this creates two columns
\hline       
Spectral res. (=mm-wave resonator quality factor) ($Q_r = R$) & 700\\ 
\hline
Optical bandwidth per detector ($\delta \nu$)& $0.3-0.7\,$GHz (e.g. 0.4) \\
\hline
Estimated system optical efficiency & 0.25 - 0.5 $\times$ 1 pol\\
\hline
Photon occupation number at the detector ($n$) & 0.5 - 5 (e.g. 1)\\
\hline
Optical power per detector (W) & 0.5 - 3.0 $\times 10^-13$ \\
\hline
Photon NEP at the detector (${\rm W\, Hz}^{-1/2} $)& 2-4 $\times 10^{-18}$\\
\hline
Operating temperature ($T_{\rm op}$ )& $250\,$mK\\
\hline
TiN transition temp ($T_c$ )& $2\,$K\\
\hline
KID resonator quality factor& $10^5$\\
\hline
TiN film thickness ($t$) and line width ($w$)& $10\,$nm, $1\,\mu$m\\
\hline
Inductor Volume & $135\mu{\rm m}^3$\\
\hline
Quasiparticle recombination time ($\tau$ )& $30\,\mu$s\\
\hline
Photo-produced quasiparticle number ($N_{qp} $)& 1-10 $\times 10^4$\\
\hline
Photo-produced quasiparticle density ($n_{qp} $)& 80-800 $\mu{\rm m}^{-3}$\\
\hline
Thermal quasiparticle density ($n_{qp}^{th} $)& $\ge 4$ $\mu{\rm m}^{-3}$\\
\hline
%\rule[-1ex]{0pt}{3.5ex}  
\end{tabular}
\end{center}
\end{table} 

The fundamental limit to the sensitivity of the SuperSpec design will be \ac{TLS} noise.  
This source of excess noise was detected early in in the development of \acp{KID}\cite{day03} and is now known to arise from fluctuations of the resonator capacitance due to the presence of microscopic \ac{TLS}
fluctuators in amorphous dielectrics\cite{gao07, kumar08, gao08,noroozian09}.
The noise does not arise in the kinetic inductance detecting element itself, so it is possible
to engineer the device to bring the TLS noise well below the fundamental photon noise. For
SuperSpec, our approach is to bring the readout frequencies down from a few GHz to a few hundred MHz.
This strategy can be understood by examining the condition on the spectral density of TLS
fractional frequency noise $S_{\rm TLS}$ for achieving photon-noise limited operation, \cite{zmuidzinas12}

\begin{equation}
S_{\rm TLS} < A \frac{\beta^2}{4 Q^2_{\rm qp}}\frac{1+n_0}{n_0 \delta \nu}
\end{equation}

\noindent Here  $n_0 \sim 1$ is the photon occupation number, $\delta \nu \sim 400\,$MHz is the spectral resolution,
$Q_{\rm qp}$ is the internal quality factor due to quasiparticle dissipation, $A$ is a dimensionless factor of order unity, 
and $\beta$ is the ratio between the frequency and dissipation response of the resonator discussed previously.
For a readout frequency of 200 MHz and operating temperature of 250 mK, we have $\beta \approx 26$. We thus
require $S_{\rm TLS} \lesssim 2 \times 10^{-17} {\rm Hz}^{-1}$. Recent measurements by our group indicate that 
for interdigitated capacitors on bare silicon, this should be a straightforward requirement to meet. \cite{mckenney12}

Each KID is coupled to the readout feedline using a small interdigitated capacitor, one side of which is patterned in the TiN material while the other is patterned in the Nb feature layer.  Because the coincidence of a short across the coupling capacitor {\em in addition} to a pinhole short to the ground plane in the inductor of a single KID would effectively disable the entire feedline circuit, we've chosen to use $2\mu$m features and $2\mu$m gaps for all of the coupling capacitor IDCs.  

Since the inductor portion of the KID is beneath a dielectric layer and ground plane, the capacitance between the inductor and the ground plane dominates the current return-path to ground.  For the dimensions shown here, we expect this capacitance, $C_g$, to be roughly $0.5\,$pF.  So long as this value is small compared to the resonator capacitance, the loss mechanisms present in this parallel plate capacitor can be neglected.  %The equivalent circuit is shown in figure \ref{fig:circuit}.  
The coupling capacitor is chosen for a target value of $Q_c$, given by

\begin{equation}
Q_c = \frac{8C}{\omega_0 C_e^2 Z_0}
\end{equation}

\noindent where the effective coupling capacitance, $C_e = (C_c\,C_g)/(C_c+C_g)$, approaches $C_c$ in the case where $C \gg C_g \gg C_c$.  

Each resonator is connected via its coupling capacitor to a $50\,\Omega$ \ac{CPW} readout feedline. This feedline has a $7\,\mu$m wide center conductor made from the Nb feature layer, and a $4\mu$m gap.  The CPW ground plane is made from the top-layer Nb ground plane, and is continuous across most of the surface area of the chip.  CPW ground straps, consisting of $5\mu$m wide bridges of the ground plane over the top of the center conductor, with intact dielectric between the two, are placed approximately every $250\,\mu$m along the feedline.  The input and output of the readout line consist of a tapered, fixed-impedance transition to wirebond pads that connect to a matching CPW line on the readout package.

To maximize the likelihood of achieving our target frequency spread, all of the inductors in the SuperSpec prototype are designed to have the same properties.  The small differences in inductor length associated with the variable gap between the inductor and fixed-width tines is compensated by an adjustable length of vertical inductive line connecting the capacitor and the meander.  To determine the appropriate capacitor for each \ac{KID} resonator, a series of SONNET simulations are run on models \ac{KID}s spanning a range of discrete capacitor sizes, and the results were fit to a semi-analytic model which was then used to determine the capacitor length corresponding to a target frequency.  

\section{Prototype test device}
\begin{figure}[hbt!]
   \begin{center}
   \begin{tabular}{c}
   \includegraphics[width=.7\textwidth]{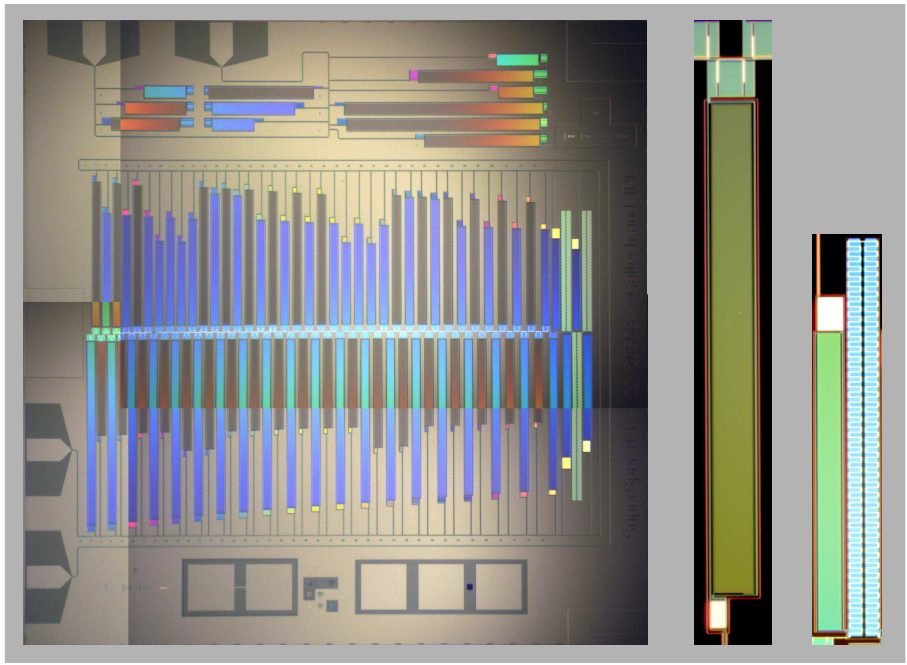}
   \end{tabular}
   \end{center}
   \caption[example] 
   { \label{fig:photo}
     {\bf (left)} A mosaic image made from several microscope photographs of the first SuperSpec prototype device.  The die is a square with side length $1.08\,$cm.  The (electrically disconnected) mm-wave feedline begins in the center left of the image and couples to the main row of mm-wave resonators and associated \acp{KID} before reaching the terminating meander at the center right.  The main readout line starts at the bottom left and connects to the coupling capacitor of each main line \ac{KID}, then returns to a second pair of CPW bond pads at the bottom left.  A second feedline couples to a sparse array of 12 test devices at the top of the image.  Additional test structures for measuring resistivity and step coverage reliability are distributed along the bottom of the die. {\bf (center)} An enlarged image showing a single \ac{KID} resonator coupled to the mm-wave feedline. {\bf (right)} An enlarged image of a portion of the termination meander.}
   \end{figure}

The first SuperSpec test device consists of all of the mm-wave device and \ac{KID} features; however, it does not include the horn-coupling hardware and cannot receive optical power. The same mask set should be compatible with future horn and waveguide-probe designs.  The device consists of a $1.16\,{\rm cm}^2$ die fabricated on a stock $100\,$mm-diameter silicon wafer.  The process includes 5 depositions (TiN, a protective \siotwo  layer over the TiN, feature Nb, \ac{sinx} dielectric, and Nb ground plane), and four lithographic steps, of which only two include critical $\mu$m-scale features and alignment.  The final, horn-coupled design will be fabricated on a \ac{SOI} wafer with a $25\mu$m device layer, and will include two additional deep reactive ion etch (DRIE) steps to define the probe and the die outline.  A photograph of the prototype die is shown in figure~\ref{fig:photo}.

The prototype device includes three types of mm-wave features: 74 tuned mm-wave filters, 3 in-line broad-band detectors, and a terminating absorber.  The 74 tuned filters span the $200-300\,$GHz band.  These include both isolated, individual filters and groups of five with a range of oversampling factors ($\Sigma$).  Design values of $Q_{\rm det}$ and $Q_{\rm feed}$ independently sample the range from 600 to 2800, and include the demonstration instrument goal of $2\mathcal{R}=1400$.  The in-line broad-band detectors consist of a short ($< \lambda_{\rm min}/4$) feedline meander in close proximity to a \ac{TiN} absorber similar to that used in the tuned filters.  Sonnet simulations predict roughly $0.5\%$ absorption across the full optical band with a slowly varying frequency dependence for these test devices.  The terminating absorber consists of several cm of meandered feedline surrounded by \ac{TiN} meander with a $1\,\mu$m separation, and is designed to absorb any radiation which arrives at the end of the feed and reduce reflections to $<20$dB.  Four long segments of the terminating absorber are used as the inductors for an addition set of broad-band absorber \acp{KID}. These are designed with readout frequencies well outside of the band of the typical \acp{KID} in order to avoid frequency collisions arising from the imperfect simulation of inductors with different geometries.

The mm-wave frequencies of the main readout line devices are arranged in descending, monotonic order.  The readout frequencies are arranged so as to minimize the potential for electromagnetic coupling between devices and to allow for the unambiguous identification of individual channels. \cite{noroozian12} The band is broken into seven discrete sub-bands, as shown in figure~\ref{fig:f_vs_f}, and neighboring pixels are drawn from different sub-bands such that each pixel is separated from its four nearest neighbors by at least $1/4$ of the full readout bandwidth.  The three smaller sub-bands all correspond to pixels with a the same design values of $Q^{mm}_r$, which will facilitate laboratory tests conducted with a broad-band optical source.

All the \acp{KID} which surround the mm-wave line are connected to the same CPW readout feedline.  In addition to this main readout line, the prototype device also includes a sparse array of 12 test resonators.  These \acp{KID} are similar to the standard design, but include a range of inductor sizes and both $1\mu$m and $2\mu$m IDCs, arranged so that each combination is repeated across a range of roughly 1.5 in readout frequency.  The test devices also include a design variation in which the ground plane and dielectric is entirely removed from the area surrounding the inductor and an explicit parallel plate capacitor instead provides current return to ground.  

\section{Initial test results}

Following fabrication, a group of seven dies were measured for room-temperature resistance and shorts to ground.  All but one had similar values and closely matched expectations.  One of these dies was then mounted in a light-tight sample box and tested cryogenically.  The sample was connected to the cryostat input through a $40\,$dB cold attenuator, and the output connected to a cryogenic silicon-germanium amplifier with approximately $35\,$dB gain and then to the cryostat output ports.  The stage was cooled with a dilution fridge to a base temperature of $20\,$mK.  The device was then connected to a \ac{VNA} though an additional $90\,$dB of warm attenuation on the input and $\sim 90\,$dB of warm amplification on the output.  We see evidence of the onset of bifurcation at a readout power of approximately $-130\,$dBm at the device input.

\begin{figure}[hbt!]
   \begin{center}
   \begin{tabular}{c}

   \includegraphics[width=.7\textwidth]{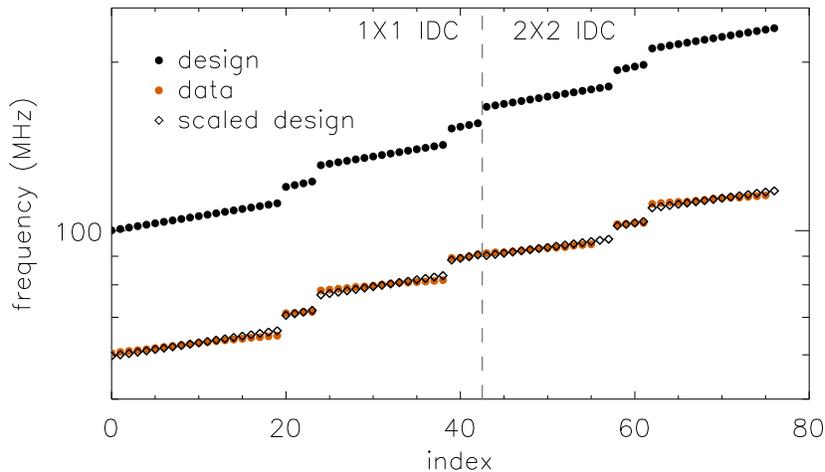}
   \end{tabular}
   \end{center}
   \caption[example] 
   { \label{fig:f_vs_f}
     A comparison of the design frequencies and observed frequencies for the prototype devices.  The vertical line separates the devices which have $1\mu$m IDC features from those with $2\mu$m IDC features.  The design values (black dots) can be be roughly fit to the observed resonances (red dots) by applying a linear scaling (black diamonds).}
   \end{figure}

Device yield for this prototype chip appears promising: resonances were seen for 74 of the 77 typical \acp{KID} on the main feedline.  Three of the four expected low-frequency, termination \acp{KID} were also seen, although it is likely the fourth landed below the $40\,$MHz cutoff frequency of the \ac{VNA}.  All of the measured devices appear to be lower in frequency than the design value by approximately 55\%.  The measured distribution of frequencies is reasonably fit by applying a single linear correction independently to all of the devices with $1\,\mu$m and $2\,\mu$m IDCs.  Comparing the data to the corrected design values, we can confidently identify the individual channels and determine that the three missing devices are all have IDCs with $2\mu$m features, as shown in figure~\ref{fig:f_vs_f}.

At present, we have not yet determined the cause of the observed frequency shift, although we expect to resolve the matter easily.  The most likely candidates are some combination of a smaller than expected TiN linewidth, a thinner than expected TiN layer, higher than expected TiN resistivity, or a lower than expected value of $T_c$.  (Mattis-Bardeen theory predicts the last three terms modify the kinetic inductance as $L_s \propto R_s / \Delta$).  A preliminary examination of the variation of resonator frequency with temperature suggests that $T_c$ may be significantly lower than the design value, most likely just below $1\,$K.  We are currently in the process of obtaining additional data to verify this.

Each resonance was measured at a fridge temperature of $20\,$mK, far below our expected operating temperature.  Under these conditions, $Q_c$ should remain unchanged, but $Q_i$ will be determined by a combination of the residual quasiparticle density in the TiN and losses in the circuit.  In this case, the dominant source of loss should be the lossy dielectric through which the \ac{KID} inductor is weakly capacitivy coupled to the ground plane.  
After corrections to account for amplifier gain variation and cable delay, a fit was performed to the complex resonator response, given by
\begin{equation}
S21(f)=a(1 - \frac{Q_r/Q_c e^{j \phi_0}}{1 + 2 jQ_r((f-f_0)/f_0)})
\end{equation}

\noindent where $a$ and $\phi_0$ are an arbitrary amplitude and phase, $f$ is the measured frequency, and $f_0$ is the resonant frequency.

\begin{figure}[hbt!]
   \begin{center}
   \begin{tabular}{c}
   \includegraphics[width=.7\textwidth]{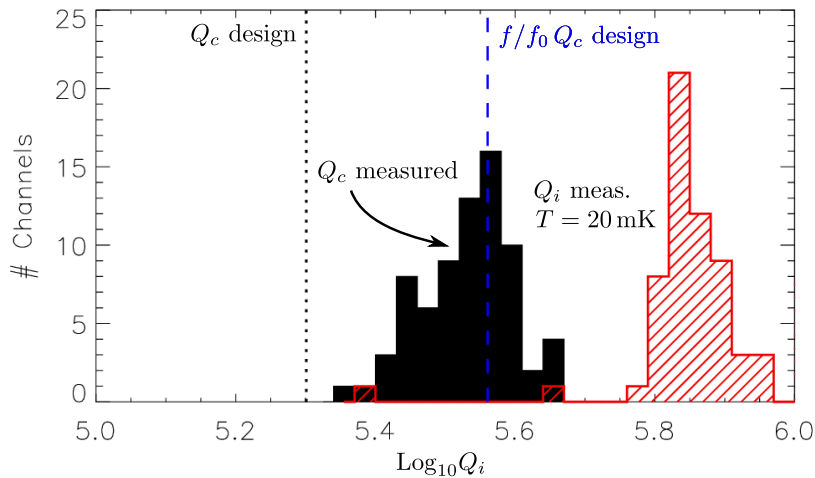}
   \end{tabular}
   \end{center}
   \caption[example] 
   { \label{fig:q_vs_q}
     The black filled histogram shows the measured values of $Q_c$ for the prototype device.  Black dotted line is the target value of $Q_c$, which becomes the blue dashed line after scaling by observed resonator frequency.  The red hashed histogram shows measured values of $Q_i$ well below the design operating temperature.}
   \end{figure}

A histogram of the fitted values of $Q_c$ and $Q_i$ are shown in figure \ref{fig:q_vs_q}.  Measured values of $\log Q_c$ are tightly clustered, with a standard deviation of $0.06\,$dex.  The mean value of $<Q_c>=3.6\times 10^5$ is higher than the design value; however, if we assume that the coupling capacitor value is correct and scale the design value to account for the frequency offset, we accurately predict the measured value.  The large measured values of $Q_i$ indicate that there are no unexpected loss mechanisms in the circuit, and that with an appropriate stage temperature and loading we should be able to reach our target value of $Q_r$.

Additional laboratory tests of the device described here are currently in progress, and we expect soon to report on both the fractional frequency noise and $Q_i$ as a function of temperature, from which responsivity and (with some assumptions about coupling efficiency) an optical NET can be estimated.

\section{conclusion}

We have presented the initial design of an ultra-compact spectrometer-on-a-chip for millimeter and submillimeter wavelength astronomy.  Detailed simulations have been used to design a filter bank spectrometer consisting of planar, lithographed superconducting transmission line resonators coupled to \ac{KID} detectors, and to verify the design is robust to realistic fabrication and design tolerances.  Lumped-element, TiN \acp{KID} have been designed which couple well to the mm-wave radiation, are optimized for the expected loading and stage temperature, and will accomodat the multiplexing of $\sim 600$ channels per octave at readout frequencies of hundreds of MHz.

An initial, dark prototype array consisting of 77 \acp{KID} and additional test structures has been fabricated.  Early results from the first cryogenic tests of this device show a very high yield, consistant resonator frequencies, tightly clustered coupling-$Q$s consistant with the design goals, and adequately high low-temperature dissipation-$Q$s.  Testing of these devices is ongoing.

Our collaboration has designed and fabricated a machined, metal feedhorn optimized for the lower SuperSpec band.  A broad-band waveguide transition to the microstrip on our devices has been simulated and is currently being fabricated.  The design of an optical prototype, which will be identical to the dark test device except for the addition of mm-wave probe features is in progress.  In the coming year we expect to demonstrate a pair of observation-grade, $600$ channel, $\mathcal{R}=700$ spectrometer pixels, one operating in the 1-mm atmospheric window and the other covering the 650 and 850 micron bands.  This instrument will serve as a pathfinder for future multi-pixels cameras, including a proposed CCAT instrument currently called {\em X-Spec}, which will consist of hundreds of SuperSpec pixels.

\acknowledgments     %>>>> equivalent to \section*{ACKNOWLEDGMENTS}       
This project is supported by NASA Astrophysics Research and Analysis (APRA) grant no. 399131.02.06.03.43.  E. Shirokoff, C. McKenney, and L. J. Swenson acknowledge support from the W. M. Keck Institute for Space Studies.  M. I. Hollister, L. J. Swenson, and T. Reck acknowledge support from the NASA Postdoctoral Programme.  P. S. Barry acknowledges the continuing support from the Science and Technology Facilities Council Ph.D studentship programme and grant programmes ST/G002711/1 and ST/J001449/1.  Device fabrication was performed the JPL Microdevices Laboratory.

%%%%%%%%%%%%%%%%%%%%%%%%%%%%%%%%%%%%%%%%%%%%%%%%%%%%%%%%%%%%%
%%%%% References %%%%%

\bibliography{superspec01}   %>>>> bibliography data in report.bib
\bibliographystyle{spiebib}   %>>>> makes bibtex use spiebib.bst

\end{document}